\documentclass[12pt,preprint]{aastex}

\shortauthors{Shi et al.}

\begin{document}

\title{9.7 {$\mu$}m Silicate Features in AGNs: New Insights into Unification Models}

\author{Y. Shi\altaffilmark{1}, G. H. Rieke\altaffilmark{1}, D. C. Hines\altaffilmark{2}, 
V. Gorjian\altaffilmark{3}, M. W. Werner\altaffilmark{3}, 
 K. Cleary\altaffilmark{3}, F. J. Low\altaffilmark{1}, P. S. Smith\altaffilmark{1}, J. Bouwman\altaffilmark{4}}

\altaffiltext{1}{Steward Observatory, University of Arizona, 933 N Cherry Ave, Tucson, AZ 85721, USA}
\altaffiltext{2}{Space Science Institue 4750 Walnut Street, Suite 205, Boulder, Colorado 80301}
\altaffiltext{3}{Jet Propulsion Laboratory, MC 169-327, California Institute of Technology, 
4800 Oak Grove Drive, Pasadena, CA 91109}
\altaffiltext{4}{Max-Planck-Institut fu$\:$r Astronomie, D-69117
Heidelberg, Germany}

\begin{abstract}

We describe observations of 9.7 {$\mu$}m silicate features in 97 AGNs,
exhibiting a  wide range of AGN  types and of  X-ray extinction toward
the central nuclei.  We find that the strength of the silicate feature
correlates with the HI column density estimated from fitting the X-ray
data, such that  low HI columns correspond to  silicate emission while
high  columns  correspond to  silicate  absorption.   The behavior  is
generally consistent with unification models where the large diversity
in AGN properties is  caused by viewing-angle-dependent obscuration of
the nucleus.   Radio-loud AGNs and radio-quiet  quasars follow roughly
the correlation  between HI columns  and the strength of  the silicate
feature defined  by Seyfert galaxies.   The agreement among  AGN types
suggests a high-level unification with similar characteristics for the
structure of the obscuring  material.  We demonstrate the implications
for  unification models  qualitatively with  a conceptual  disk model.
The model includes  an inner accretion disk ($<$ 0.1  pc in radius), a
middle disk (0.1-10  pc in radius) with a  dense diffuse component and
with embedded  denser clouds, and an  outer clumpy disk  (10-300 pc in
radius).

\end{abstract}


\section{Introduction} 
The arrangement of material  around the central supermassive blackhole
(SMBH)  in active  galactic nuclei  (AGNs) is  crucial  in unification
schemes, where the  active nuclei classified as type  2 arise from the
obscuration  of  broad  optical  emission  lines  in  type  1  objects
\citep{Antonucci93}.  The structure  of the circumnuclear material may
be critical to  the growth of the SMBH and  may influence the feedback
from the  AGN and its  effects on the  formation and evolution  of the
host galaxies.   The infrared (IR) spectral  energy distribution (SED)
arises from the heating of  surrounding material, and its shape should
indicate  how the  material is  organized.  There  are at  least three
different  possibilities:  compact  \citep{Pier92,  Pier93},  extended
\citep{Granato94,     Granato97}    and     cloudy    \citep{Krolik86,
Rowan-Robinson95, Nenkova02}  structure models  all fit the  data well
(assuming an extra component of  star formation in the compact model).
The silicate emission and  absorption features now being observed with
{\it Spitzer} (e.g.,  Weedman et al. 2005) can be  used to probe these
possibilities further.

In   this   paper,   we   present  $Spitzer$   Infrared   Spectrograph
\citep[IRS;][]{Houck04} observations  of 9.7 $\mu$m  silicate features
in    93    AGNs,    including    examples   from    the    literature
\citep{Siebenmorgen05,  Hao05, Sturm05,  Ogle06}.  We  complement them
with  groundbased   measurements  for  4   more  \citep{Roche91}.   By
combining these observations with X-ray  data that explore the line of
sight toward  the central  engine, we can  gain new insights  into the
unification scheme.

\section{Sample and Data Reduction}

Our sample (see  on-line Table 1) includes a variety  of types of AGN:
radio-loud  quasars, FR  II radio  galaxies, Seyfert  1 and  Seyfert 2
galaxies,   broad  absorption-line  quasars   (BALQs),  low-ionization
nuclear emission-line region (LINER) galaxies, and non-BAL radio-quiet
quasars including optically selected Palomar-Green (PG) quasars and IR
quasars  as selected  by the  Two-Micron All  Sky Survey  (2MASS) from
\citet{Smith02}.  Except for the  objects from the literature, quasars
of all types and FR II radio galaxies are the most luminous objects in
their parent samples (which include  upper limits in redshift in their
definitions) while Seyfert 1 and Seyfert 2 galaxies are derived to have
high  brightness  and a broad   range  in  HI  column  density  from
\citet{Turner89}  and  \citet{Risaliti99},  respectively.   The  whole
sample spans a range of  HI column density from 10$^{20}$ cm$^{-2}$
up to 10$^{25}$  cm$^{-2}$. We focus on the sample  of 85 objects with
available  HI  column densities  in  this  study,  although all  97
objects have  available silicate  data.  X-ray spectra  were retrieved
from the Chandra  data archive for 8 sources.   The data reduction and
the measurement  of column densities through power-law  fits are as
described in \citet{Shi05}.  The column densities of most of the 2MASS
QSOs  are based  on  the X-ray  hardness  ratio from  \citet{Wilkes02}
assuming an intrinsic power law  X-ray spectrum with a photon index of
1.7. The  associated  uncertainty  is  estimated  as  a  factor  of  3
corresponding to a change of  $\sim$1 in the photon index.  The column
density for the  remaining sources is obtained from  the literature as
shown in the on-line Table 1.

The IRS spectra presented for the first time in this paper, except for
those  of  the Seyfert  galaxies,  were  obtained  using the  standard
staring  mode.  The  intermediate  products of  the $Spitzer$  Science
Center (SSC) pipeline S11.0.2 were processed within the SMART software
package  \citep{Higdon04}.    The  background  was   subtracted  using
associated spectra  from the  two nodded, off-source  positions.  This
also subtracts  stray light contamination from  the peak-up apertures,
and adjusts pixels with anomalous dark current.  Pixels flagged by the
SSC pipeline as  ``bad'' were replaced with a  value interpolated from
an 8-pixel perimeter surrounding  the suspect pixel.  The spectra were
extracted using  a 6.0  pixel fixed-width aperture  for the  short low
module  (SL),   and  5.0  pixels   for  the  long  low   module  (LL).  
The  spectra were calibrated using  a spectral response
function derived from IRS spectra  and Kurucz stellar models for a set
of   16  Sun-like   stars  that   exhibit:  1)   high  signal-to-noise
observations, 2)  no residual instrumental artifacts, and  3) no signs
of IR excess.   The absolute flux density scale  is tied to calibrator
stars observed by the IRS instrument team and referenced to calibrated
stellar models provided by the SSC \citep[see also][]{Bouwman06}.  The
uncertainties in the final  calibration are dominated by random noise.
The relative flux calibration across the spectrum is accurate to $\sim
1-2\%$ \citep{Hines06, Bouwman06}.

The  data reduction was  slightly different  for the  Seyfert galaxies
first presented in this paper.  The whole Seyfert sample consists of a
mixture of  point-like and extended  sources.  They are  at relatively
low redshift ($z\sim$0.01) and are  lower luminosity AGNs, so care was
taken  to minimize  the contribution  of  the host  galaxies. For  the
extended  sources, a  fixed-width  column extraction  that is  narrow
enough  to  exclude  most  of   the  extended  source  would  be  most
appropriate.   However,  narrow  fixed-width  extraction  windows
introduced artifacts  into the extracted spectra.   Instead, a narrow
expanding-aperture extraction was performed on all sources, point-like
or extended.  This  was deemed to be an  acceptable compromise between
rejecting  the  extended source component  and  introducing  artifacts into  the
spectrum.  The  spectra were  extracted using an  expanding extraction
aperture  defined to be 4 pixels wide at  the wavelengths of 6
$\mu$m  and 12  $\mu$m  for the  second  and the  first  order of  SL,
respectively, and  of 16 $\mu$m and  27 $\mu$m for the  second and the
first order of  LL, respectively. At the wavelength  of 10 $\mu$m, the
extraction aperture  is around 3.3 pixels.  A  detailed description of
the data reduction is given by \citet{Gorjian06}.

\begin{figure}
\epsscale{1.}
\plotone{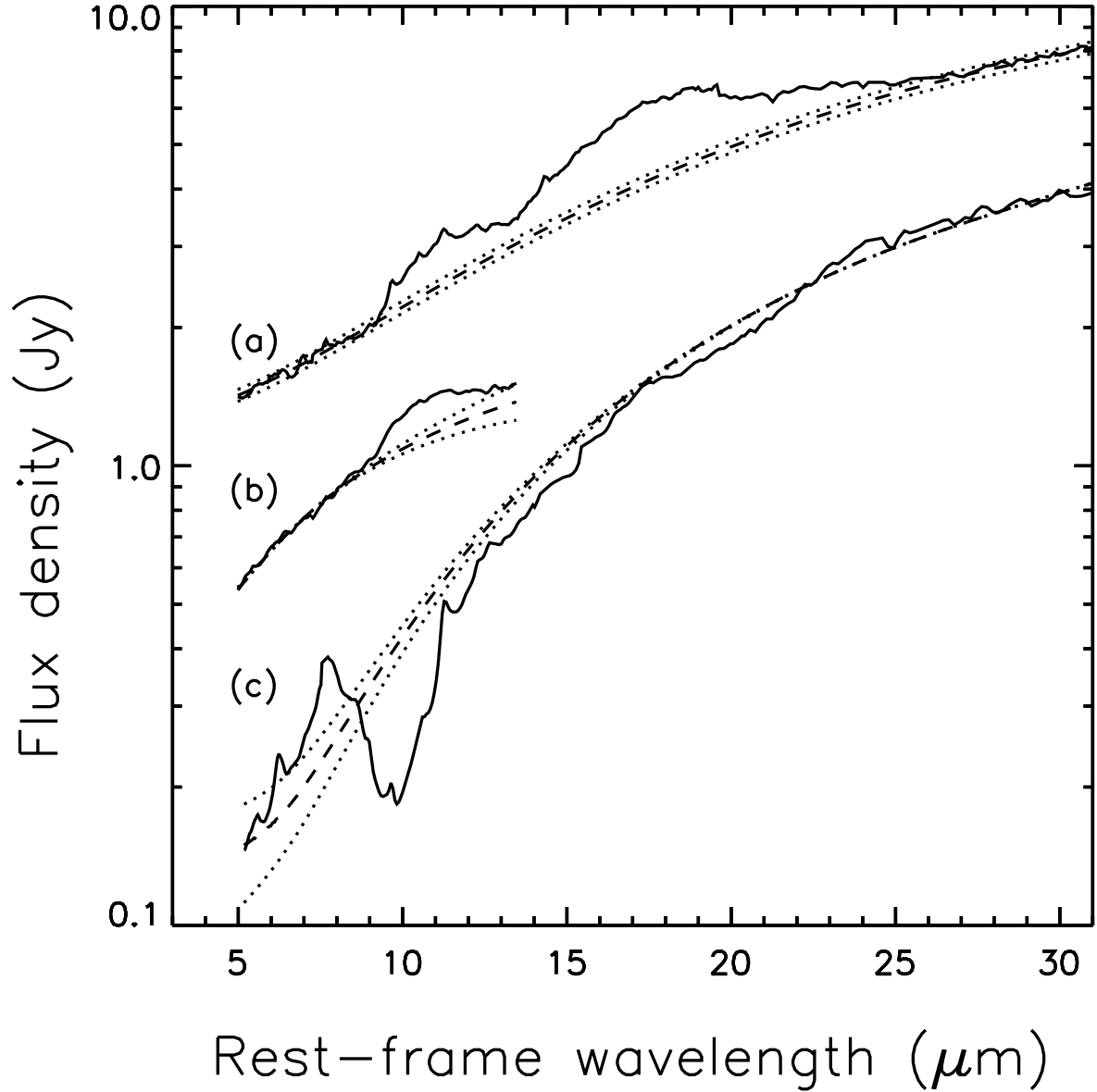} 
\caption{ \label{spec} IRS spectra of 3 objects with fits. The dashed line is the fitted
continuum and the dotted line is the estimated uncertainty in the continuum fitting. The scaling factors 
for (a) MCG-2-58-22, (b) Fairall9 and (c) NGC4388, are 30, 5 and 1.5, respectively. }
\end{figure}

The  silicate feature  strengths  were estimated  as follows.   Narrow
emission  lines were  removed  from  the spectra  and  they were  then
smoothed to a resolution of  0.1 $\mu$m.  The continuum was defined by
using a spline interpolation between the  blue and red ends of the IRS
spectral range.  We defined (all in rest-frame wavelengths) the blue
end as 5-7.5 $\mu$m, while the red end was defined as 13-14 $\mu$m for
the 13  objects observed only  with the SL  module.  For six  BALQs at
higher redshifts,  the red end was  chosen to be  10.5-13 $\mu$m.  For
the remainder of the sample, the red spectral end was defined as 25-30
$\mu$m.  For  nine objects with  the red end contaminated  by silicate
emission  (Figure~\ref{spec}(b)),  the   continuum  at  this  end  was
estimated to be  below that observed by $<$20\%  at the reddest point,
based  on the spectra  of objects  without contamination.   For twelve
objects  with the  blue  end  contaminated by  the  strong 6-8  $\mu$m
aromatic  feature   (Figure~\ref{spec}(c)),  the  continuum   in  this
bandpass was given by two segments  between 5 and 5.5 $\mu$m where the
contamination is  negligible.  The fit was  judged to be  good when it
matched the  flux at both ends  within the noise and  the curvature of
the  continuum varied gradually  from one  end to  the other  over the
silicate region.  The uncertainty in continuum fitting was obtained by
adjusting  the flux  of the  continuum up  and down  over  the fitting
spectral range.   For the objects  with continuum not  contaminated by
either aromatic or silicate features as shown in Figure~\ref{spec}(a),
we  adjusted the fit  until the  continuum was  just above  (for upper
error) or  below (for  lower error) the  observed flux over  the whole
spectral fitting  range.  To estimate the uncertainty  for the objects
with    red    end    contaminated    by    the    silicate    feature
(Figure~\ref{spec}(b)), we  first fixed the blue end  and adjusted the
red  end  continuum  up  to   the  red  end  point  that  matched  the
upper-envelope of the underlying continuum  plus noise and down by the
same amount  as an estimate  of the errors  of the continuum  fit.  To
account for the effects of noise in the blue end, we added another 4\%
uncertainty (the  mean value for objects without  aromatic or silicate
feature contamination).  For the objects with blue end contaminated by
the aromatic feature (Figure~\ref{spec}(c)),  we fixed the red end and
adjusted the  blue end to get  the uncertainty in a  similar way where
the upper-limit  of the underlying continuum  is the flux  at 7 $\mu$m
(the median wavelength between the  peaks of the 6.2 $\mu$m and 7.7 $\mu$m
aromatic  features).   Another  4\%  uncertainty  was  added  to  this
uncertainty to account for the effects of noise in the red end.  There
are no objects  with red end contaminated by  the silicate feature and
blue end contaminated by the aromatic features.

The strength of the 9.7 $\mu$m silicate feature is defined as ($F_{\rm
f}-F_{\rm c})/F_{\rm  c}$, where $F_{\rm  f}$ and $F_{\rm c}$  are the
observed   flux  density  and   underlying  continuum   flux  density,
respectively,  at  the  peak   (for  emission)  or  the  minimum  (for
absorption)  of the silicate  feature.  The  feature strength  in this
definition is a  direct measure of the optical  depth for the silicate
absorption. Figure~\ref{all_spec} shows  the IRS spectra presented for
the first time in this study in order of HI column density.  No ice
or   hydrocarbon  absorption   is   found  in   Figure~\ref{all_spec}.
Especially  at 5-8  $\mu$m, where  there  is no  contamination of  the
silicate feature, the spectra are  well described by power laws except
for those  with aromatic emission.  Therefore, once  the continuum had
been fitted, we could calculate the feature strengths unambiguously.

\section{Results}

Figure~\ref{NH_Silicate} shows the strength of the 9.7 $\mu$m silicate
feature as  a function of HI  column density. The  right y-axis of
Figure~\ref{NH_Silicate}   shows  the  IR-absorbing   column  density
estimated   from  the   silicate  absorption   feature   by  assuming
$\tau_{9.7{\mu}m}={\ln}(F_{\rm      c}/F_{\rm      f})$,      $A_{\rm
v}/A_{9.7{\mu}m}$=19        \citep{Roche85}        and        $A_{\rm
v}/N_{H}$=0.62$\times10^{-21}$cm$^{-2}$     \citep{Savage79}.     The
silicate feature  varies from emission  (+) to absorption (-)  as the
X-ray  spectra  become more  heavily  obscured.   The  trend is  also
demonstrated  by the  composite spectra  in different  bins  of HI
column as  shown in Figure~\ref{CP_spec}. Since  the silicate feature
is broad,  the composite  spectra are the  geometric mean  spectra to
conserve the global spectral shape \citep[e.g.][]{VandenBerk01}.  The
relationship between  HI column  and silicate feature  behavior is
generally consistent  with the AGN unification  scheme where material
surrounding  the central SMBH  obscures both  the X-ray  and silicate
emissions.

The correlation  (black lines in  Figure~\ref{NH_Silicate}) is defined
by all  types of AGN. To  investigate the behavior  for individual AGN
types, we unified Seyfert 1 and  Seyfert 2 galaxies as one type of AGN
(Seyfert  galaxies). Similarly,  FR II  radio galaxies  and radio-loud
QSOs are  classified as  radio-loud AGNs while  PG and 2MASS  QSOs are
classified as radio-quiet  QSOs.  Table 2 shows the  linear fits to
these AGN types where the limits  to the HI column density are treated
as  detections during  the fitting.   The fits  for  different AGN
types  are  characterised  by  large  uncertainty  and  are  generally
consistent with  each other within the  uncertainty.  Radio-quiet QSOs
(2MASS  and  PG  QSOs)  exhibit   a  small  (less  than  two  standard
deviations)  deviation from  other  two  types of  AGN.   As shown  in
Figure~\ref{NH_Silicate},  this deviation arises  because the  PG QSOs
with low HI columns have larger silicate strengths.

To  quantify  the  discrepancy  between  different types  of  AGN,  we
examined the probability  that the other AGN types  follow the Seyfert
correlation by  producing a Monte Carlo  theoretical distribution with
10000  data  points  using   the  Seyfert  correlation  (red  line  in
Figure~\ref{NH_Silicate}) with associated  scatter.  Table 3 shows the
result of three tests: 1.) the probability  from a K-S test that the given
AGN type  has the same  distribution as the  theoretical distribution; 2.)
the significance that two  distributions have the same mean (indicated
by  a  value  greater  than  0.05);  and 3.) the  significance  that  two
distributions  have the same  variance (indicated  by a  value greater
than  0.05).   The  high  K-S  probability for  the  Seyfert  galaxies
indicates that  the theoretical distribution is  a good representative
of   them   (as  it   should   be).   As   shown   in   Table  3   and
Figure~\ref{NH_Silicate}, radio-loud  AGNs (radio-loud QSOs  and FR II
galaxies) are consistent with  the Seyfert correlation and radio-quiet
QSOs (2MASS  and PG QSOs)  show slightly higher silicate  strengths at
low HI column compared to Seyfert galaxies.

\citet{Sturm06} recently found  that six type 2 QSOs  in the HI column
range  of 10$^{21.5}$-10$^{24}$  cm$^{-2}$  do not  show any  silicate
feature,  slightly  higher but  still  consistent  with  those of  the
Seyfert  galaxies  at the  same  range of  HI  columns.   As shown  in
Figure~\ref{NH_Silicate}, the  three BALQs and  the one LINER  in this
study deviate from the Seyfert correlation significantly. More objects
are  needed to  address whether  these  two types  follow the  Seyfert
correlation. Given  that most PG QSOs have  upperlimit measurements of
HI  columns, the  intrinsic  deviation of  radio-quiet  QSOs from  the
Seyfert  trend should  be  smaller than  we  calculate.  Although  the
intrinsic slope  of the Seyfert  correlation should be smaller  due to
the lowerlimit  measurement of  the HI column  for half the  Seyfert 2
galaxies,  the  effect on  the  comparision  is  small because  it  is
dominated by the column range  where Seyfert galaxies have detected HI
columns.  In Table 2, we also list the linear fits to the Seyferts and
the whole sample  excluding objects with limit measurements  of the HI
columns.   The  comparision between  the  new  slopes  shows that  the
consistency between different AGN types may become even better.

Therefore, except for LINERs  and BALQs, the remaining non-Seyfert AGN
types  follow  more  or   less  the  Seyfert correlation.   The rough
agreement  among  AGN  types   indicates  that  the  geometry  of  the
circumnuclear material  and our viewing angle are  the primary factors
influencing the  relation between the  silicate feature and  the X-ray
attenuation.   This  correlation is  nonetheless  curious because,  as
shown  in  Figure~\ref{NH_Silicate},   the  column  required  for  the
observed X-ray obscuration levels  is up to $\sim$10$^{25}$ cm$^{-2}$,
two  orders of  magnitude larger  than  that required  to produce  the
silicate  absorption, $\sim$ 10$^{23}$  cm$^{-2}$.  This  implies that
along many lines of sight  there must be X-ray absorbing material that
is not contributing to the silicate absorption.

The   dispersion  in  the   correlation  is   much  larger   than  the
uncertainties in  silicate-feature strength and  HI column density.
The large dispersion appears to be characteristic of all the AGN types
with large numbers of  objects (Seyfert galaxies, radio-quiet QSOs and
radio-loud  AGNs).  This is  consistent with  a common  mechanism (the
circumnuclear geometry)  regulating the correlation  and dispersion in
Figure~\ref{NH_Silicate}.    Due   to   different  physical   aperture
diameters from 500  pc to galaxy-scale in size,  the differing amounts
of extended  emission from star-forming regions may  contribute to the
scatter.   However, such  contributions must  be small.   As  shown in
Figure~\ref{all_spec},  twelve  of  85  objects have  strong  aromatic
features  and   their  silicate   features  may  be   contaminated  by
star-forming regions.   However, the dispersion of  the correlation is
almost the same for the 73 objects without aromatic features.  Another
possible  contribution  to  the  scatter  is  the  gas-to-dust  ratio.
However,  as indicated  by  Figure~\ref{NH_Silicate}, the  gas-to-dust
ratio needs  to vary nearly three  orders of magnitude  to account for
the scatter  in the correlation. Such large  variation is unreasonable
given the similar IR SEDs and IR luminosities for objects with similar
HI  column   but  different   silicate   features   as  shown   in
Figure~\ref{all_spec}, for example, NGC 4941 versus NGC 3281.

Another   characteristic   of   the   correlation  is   that   several
Compton-thick AGNs have silicate absorptions that are much weaker than
predicted.   Again, this  implies that  additional  absorbing material
must be  present in the Compton-thick  sources, but placed  so it does
not obscure the IR emission.

\section{Discussion}

We now describe a conceptual  model to explain the large difference in
the absorbing columns  for the X-rays and infrared,  but that can also
account for the correlation  in Figure~\ref{NH_Silicate}.  The goal is
to find  a geometry  for the circumnuclear  material that  can explain
Figure~\ref{NH_Silicate}  solely  in terms  of  variations in  viewing
angle.  As mentioned  in the Introduction, there are  three classes of
model that are generally successful in fitting the IR SEDs of specific
AGN  types: 1.)  a  compact disk  \citep[e.g.][]{Pier92}; 2.)   a more
extended  disk (radius of  hundreds of  pc) \citep[e.g.][]{Granato97};
and   3.)     a   disk   with    clumps   or   clouds    of   material
\citep[e.g.][]{Nenkova02}.   We now explore  whether the  ideas behind
these  three types  of model  can be  combined to  provide  a possible
explanation for  the behavior of  the silicate feature.  It  is beyond
the scope  of this paper  to compute a  quantitative model to  fit the
data.  However, by confining  our argument to combinations of features
in models  already shown to fit  other aspects of AGN  behavior, it is
likely  that our  explanation of  the silicate-X-ray  correlation will
also be compatible  with the other observations of  AGNs. Overall, our
model  requires both  a component  of material  similar to  the cloudy
model  \citep[e.g.][]{Nenkova02} and  a diffuse  component  with outer
radius  similar to the  compact disk-model  \citep[e.g.][]{Pier92} and
with  column  density  similar  to  the model  of  the  extended  disk
\citep[e.g.][]{Granato97}.

In  Figure~\ref{model}, we  show the  hypothetical disk  geometry.  We
include an inner  accretion disk (AD), and in the  same plane a middle
disk (MD) with  a diffuse component (grey) joining  to the AD.  Denser
clouds or clumps are embedded in  this MD.  It merges with a cloudy or
clumpy outer disk  (OD).  The X-ray and UV radiation  from the AD heat
the  dust in  the MD  and OD  to produce  IR emission  and  ionize the
broad-emission-line  clouds  and   the  narrow-emission-line  (NEL  in
Figure~\ref{model})  clouds, which  are  not shielded  by the  diffuse
component.  Compton-thick  X-ray obscuration  can arise either  in the
AD,  or in  the clouds  in the  MD. Irregularities  in the  AD  or the
passage of clouds through the line of sight may be responsible for the
variations   in   X-ray   obscuration   observed  toward   some   AGNs
\citep[e.g.][]{Elvis04, Risaliti05}.   However, for a  model with only
the  central disk  and the  clouds, there  is a  large  probability of
viewing  the central  X-ray emission  directly without  any extinction
even  for an  inclined disk  (along  the equatorial  plane).  This  is
inconsistent with Figure~\ref{NH_Silicate}, where the lower-envelope
of the  data distribution shows a  strong trend that  the HI column
density  increases with  the depth  of the  silicate  absorption.  The
diffuse component of the MD  is required to obscure the X-ray emission
even if the line of sight  does not cross any cloud. This component is
also  the source of  the silicate  emission.  The  silicate absorption
results from the material in the OD.

We argue that the outer edge of the MD should not extend beyond around
10 pc  and the dust within  it can have  sufficiently high temperature
\citep[$>$ 300 K;][]{Laor93} to  produce the silicate emission.  Given
the maximum silicate absorption  of 10$^{23}$cm$^{-2}$ and the maximum
HI column of  $>$10$^{25}$cm$^{-2}$, the clouds in the  MD should have
$N_{H} > 10^{23}$ cm$^{-2}$ and produce significant obscuration of the
X-ray emission for  an intercepting line of sight  but not obscure the
silicate emission significantly due to  a small covering factor. For a
given  strength of the  silicate feature  along a  line of  sight, the
large variation  in the HI  column is caused  by the variation  in the
number or column  density of clouds that the  line of sight intercepts
and  the  minimum  X-ray  obscuration  is  due  only  to  the  diffuse
component.  Based  on this concept, as  the line of  sight varies such
that the silicate strength ranges  from 0.0 to -0.8 (the minimum value
in  the  sample)  as  shown in  Figure~\ref{NH_Silicate},  the  column
density of  the diffuse component  varies from 10$^{21}$  cm$^{-2}$ to
10$^{23}$ cm$^{-2}$. The evidence for the cloudy OD is that the mid-IR
images  of  NGC  1068  at  10  $\mu$m  \citep{Jaffe04}  and  NGC  4151
\citep{Radomski03}  at  $\sim$10 and  $\sim$18  $\mu$m  show that  the
radius of any diffuse component should be smaller than 2 pc and 35 pc,
respectively. However, the dust in the OD is relatively cool and emits
any reprocessed  energy mainly at  far-IR wavelengths and thus  may be
missed for observations at mid-IR  wavelengths.  Since both the MD and
OD are unresolved in the IRS beam, the observed silicate absorption can be
provided by several  clouds in the OD and the  depth of the absorption
feature is determined by the average  number of clouds along a line of
sight. This is because a cloud in  the OD is not large enough to cover
the whole MD.

\section{Conclusions}
We report  observations of 9.7{$\mu$}m  silicate features in  97 AGNs.
The  features  vary  from  emission  to  absorption  with  increasing
HI  column density,  consistent with  unification models.
Radio-loud AGN  (radio-loud QSOs and  FR II galaxies)  and radio-quiet
QSOs  (PG  and 2MASS  QSOs)  lie  roughly  on the  Seyfert-correlation
between HI column and  silicate feature strength.  The behaviors of
LINERs and BALQs are not clear  due to the small number of objects for
these two AGN types.  The scatter in the relation is large and several
Compton-thick   AGNs   do   not   show   deep   silicate   absorption.
Qualitatively, the correlation  requires a circumnuclear disk geometry
with an  accretion disk outside  of which is  a middle disk  with high
density and  with even denser  clouds embedded (0.1-10 pc  in radius),
co-aligned  with an  outer clumpy  disk  (10-300 pc  in radius).   The
similarity of the behavior of  various types of AGN suggests that this
disk geometry may be typical for AGN in general. \\

We thank  the anonymous referee  for detailed comments. We  also thank
Roberto Maiolino  for helpful suggestions.  Support for  this work was
provided by  NASA through contract  1255094 issued by  JPL/ California
Institute of Technology.

\clearpage

\begin{deluxetable}{llllcllll}
\tablecolumns{8}
\tabletypesize{\scriptsize}
\tablewidth{0pc}
\tablecaption{Source Characteristics}
\tablehead{
\colhead{Sources}&  \colhead{z} &   \colhead{Type} &  \colhead{Strength}  &  \colhead{Reference}   & \colhead{$N_{H}^{X}$} &  \colhead{Reference} \\
 \colhead{(1)}   &\colhead{(2)} &\colhead{(3)}     &   \colhead{(4)}       & \colhead{(5)}         & \colhead{(6)}           &  \colhead{(7)}         

}
\startdata
     PG0050+124             &     0.058 &    PG     &        0.38                     & 2  &   0.03$^{+0.01}_{-0.016}$ &     10       \\
     PG0052+251             &     0.155 &    PG     &        0.33$^{+0.06}_{-0.06}$   & 0  & 0.04 $^{+0.01 }_{-0.01 }$ &    8     \\
     PG0804+761             &     0.100 &    PG     &        0.60                     & 2  &   0.03$^{+0.00}_{-0.005}$ &     18       \\
     PG0953+414             &     0.234 &    PG     &        0.40$^{+0.08}_{-0.08}$   & 0  &                $<$  0.006 &    9     \\
\enddata

\tablecomments{ Column (1):  The sources. Column  (2): Redshift. Column
(3): The types  of AGNs. $'$PG$'$: PG quasar;
$'$RLQ$'$: radio-loud quasar; $'$Sy1$'$: Seyfert  1 galaxies; $'$Sy2$'$: Seyfert 2
galaxies; $'$BALQ$'$:  broad absorption-line  quasar; $'$FRII$'$: FR  II radio
galaxies; $'$LINER$'$: low-ionization nuclear emission-line region; $'$2MQ$'$:
2MASS quasar.  Column(4):  The strength  of  the silicate  feature as defined in \S ~2.
Column (5): The references for the silicate data: (0) This work;    (1) \citet{Siebenmorgen05};    (2) \citet{Hao05};    (3)
\citet{Sturm05};  (4) \citet{Roche91}; (5) \citet{Ogle06}.  Column (6):  The intrinsic HI
column density in the unit of $10^{22}$cm$^{-2}$.  Column (7): References for the HI column
densities:  (6) This work;  (7) \citet{Donato03}; (8) \citet{Brunner97};
(9) \citet{Porquet04};    (10) \citet{Reeves00};   (11) \citet{Leighly97};
(12) \citet{Gallagher02};  (13) \citet{Mathu00}; (14) \citet{Dewangan03};
(15) \citet{Turner89};                       (16) \citet{Risaliti99};
(17) \citet{Wilkes02}; (18) \citet{Wang96}; (19) \citet{Ptak04}; (20) \citet{Gallagher99}; (21) \citet{Wilkes05} ; (22) \citet{Worrall01}
;(23) \citet{Belsole06}; (24) \citet{Isobe02} \\
The complete version of this table is in the electronic edition of the Journal.
}

\end{deluxetable}

\clearpage

\begin{deluxetable}{lccc}
\tabletypesize{\scriptsize}
\tablewidth{20pc}
\tablecaption{The linear fits to objects with different AGN types}
\tablehead{  \colhead{Type}  & \colhead{A}   & \colhead{ B  }  &   \colhead{Number} \\
             \colhead{(1)}   & \colhead{(2)} & \colhead{(3)}   &   \colhead{(4)}   } 
\startdata
Seyfert           & 2.6$\pm$0.7  &  -0.12 $\pm$ 0.03 & 38 \\
2MQ \& PG         & 5.5$\pm$1.2  &  -0.25 $\pm$ 0.06 & 28 \\
FRII \& RLQ       & 3.9$\pm$1.6  &  -0.17 $\pm$ 0.07 & 15 \\
all objects       & 3.3$\pm$0.5  &  -0.15 $\pm$ 0.02 & 85 \\
\hline
Seyfert$^{1}$     & 4.6$\pm$1.3  &  -0.21 $\pm$ 0.06 & 23 \\
all objects$^{1}$ & 5.1$\pm$0.9  &  -0.23 $\pm$ 0.04 & 61 \\
\enddata

\tablecomments{  Column(1): The AGN types. See  caption  to Figure
3. Column(2)  and   Column(3): Parameters of the  linear   fit  $S   =   A  +
B*\log(N_{H}^{X})$ where $S$ is  the strength of the silicate feature
as defined in \S ~2 and $N_{H}^{X}$  is the HI column  density in 
the unit of cm$^{-2}$. Both parameters $A$ and $B$ are unitless due to the
definition of the silicate strength. Column(4): the total number of objects for each fit.\\
$^{1}$Fits excluding objects without measured HI columns (i.e., upper and lower limits to the HI are excluded).}
\end{deluxetable}

\clearpage

\begin{deluxetable}{lcccc}
\tabletypesize{\scriptsize}
\tablewidth{0pc}
\tablecaption{Tests of the correlation}
\tablehead{  \colhead{Type}  & \colhead{Num}   & \colhead{ K-S Prob.(\%) }  &   \colhead{Mean}   &  \colhead{Variance}   \\
             \colhead{(1)}   & \colhead{(2)}   & \colhead{(3)}                    &   \colhead{(4)}    &  \colhead{5}   } 
\startdata
Sy                       &  38   &    90   &      1.0      &      0.85  \\
2MQ   \& PG              &  28   &     2   &      0.02     &      0.35  \\ 
RLQ   \& FR II           &  15   &     10  &      0.06     &      0.40  \\
\enddata
\tablecomments{  Column(1): The  AGN type.  See caption to  Figure~\ref{NH_Silicate}.   
Column(2): the total number of a given AGN type.  Column(3): The probability
from the K-S  test that the  given AGN type  has the same distribution  as a
theoretical Monte Carlo distribution  produced by  the correlation defined  by the
Seyfert galaxies.  Column (4): The significance  that the distribution
of the given  AGN type and the theoretical  distribution have the same
mean. A  value greater than 0.05  indicates the same mean  for the two
distributions. Column  (5): The significance that  the distribution of
the  given AGN  type and  the theoretical  distribution have  the same
variance. A  value greater than  0.05 indicates the same  variance for
the two distributions.}
\end{deluxetable}

\clearpage

\clearpage

\setcounter{figure}{1}
\begin{figure}
\epsscale{.80}
\plotone{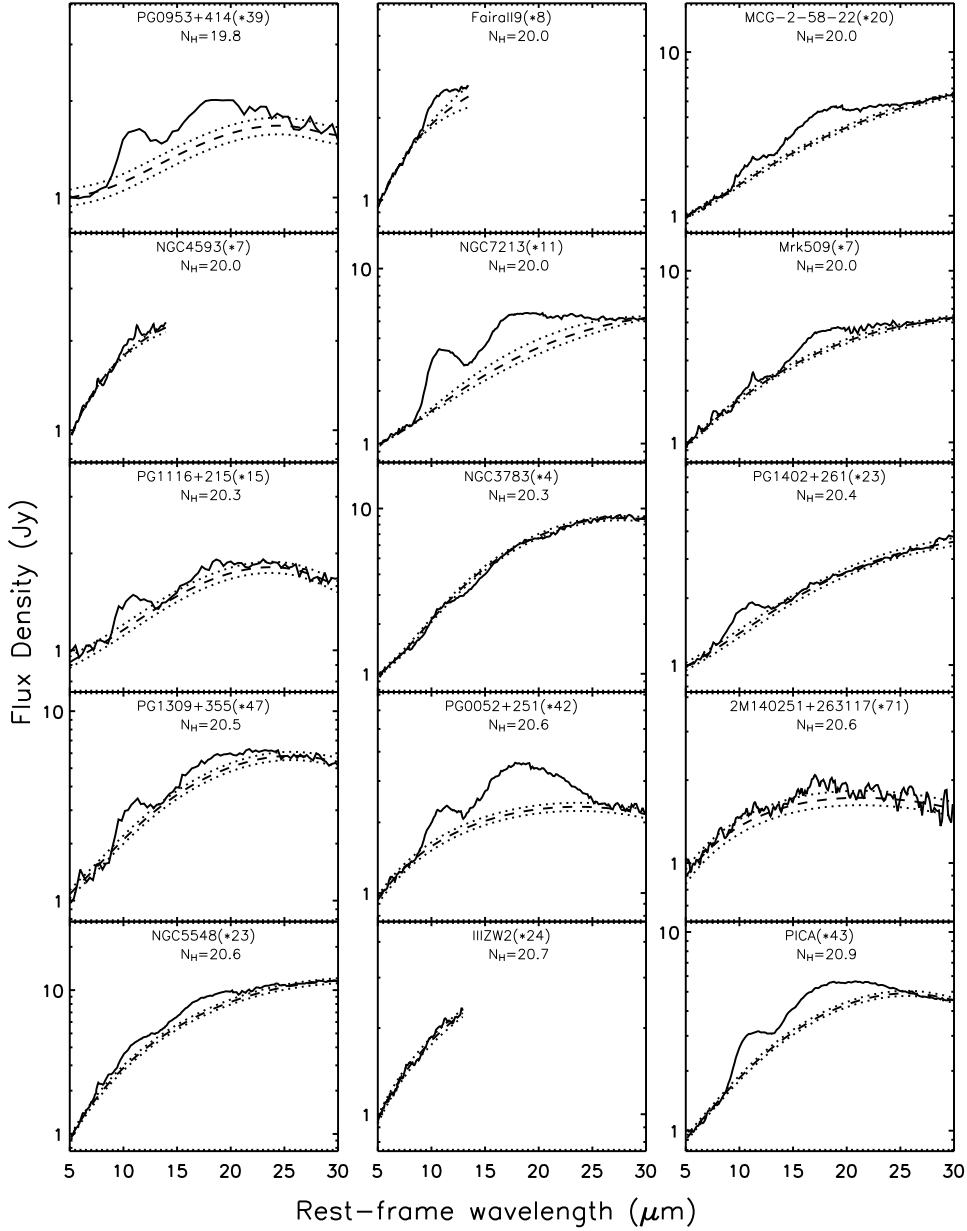}

\caption{\label{all_spec} The  IRS spectra for  objects with available
X-ray data in order of the HI column density. Narrow emission lines have been  removed from the
spectra (see text). The dashed
lines are the fitted continua  and the dotted lines show the estimated
uncertainties  in  the  continuum   fitting.  The  scaling  factor  to
normalize  the spectra  at 5  $\mu$m and  the logarithm  of  the HI
column density in units of cm$^{-2}$ are given for each object. }
\end{figure}

\setcounter{figure}{1}
\begin{figure}
\epsscale{1.0}
\plotone{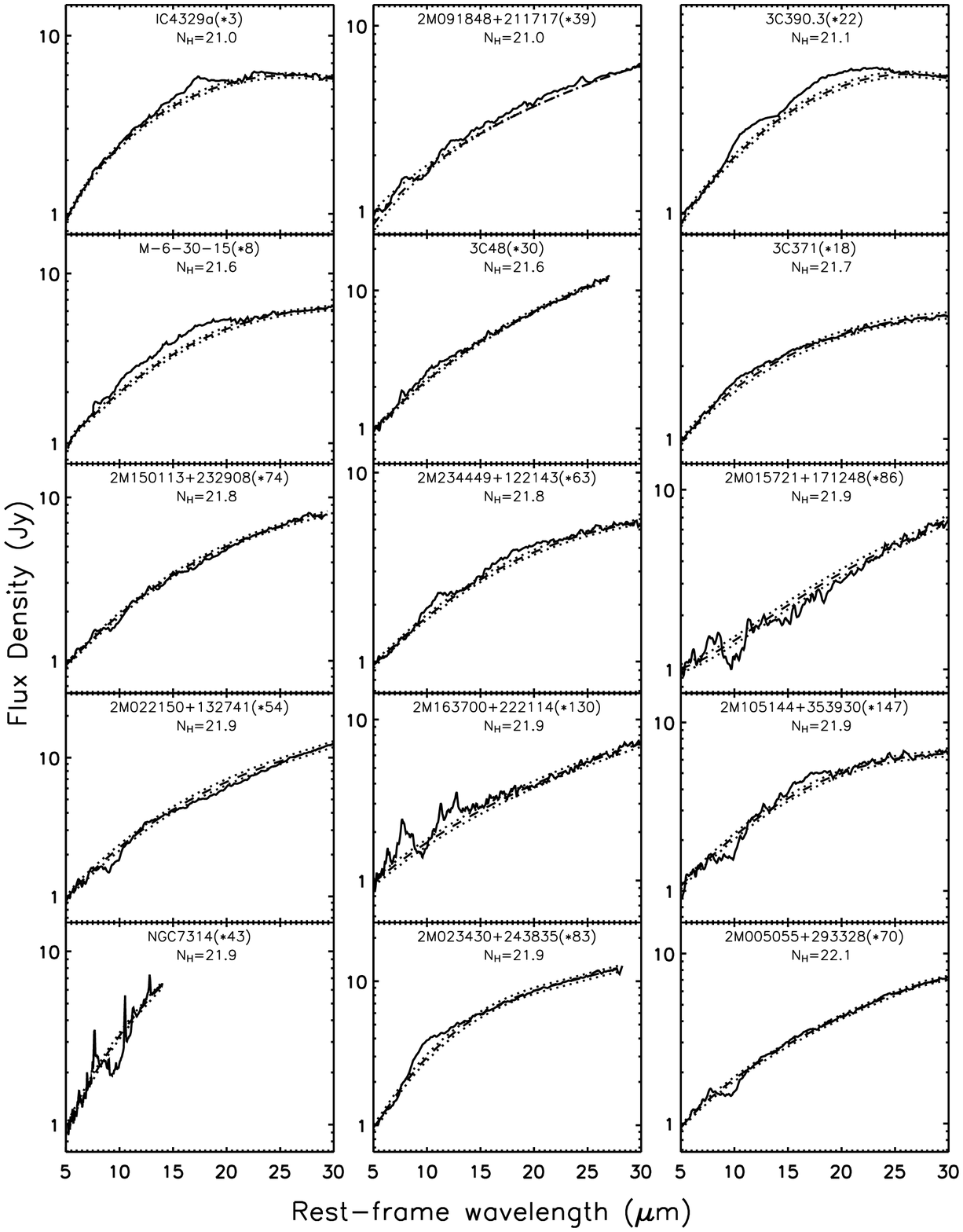}
\caption{Continued}
\end{figure}

\setcounter{figure}{1}
\begin{figure}
\epsscale{1.0}
\plotone{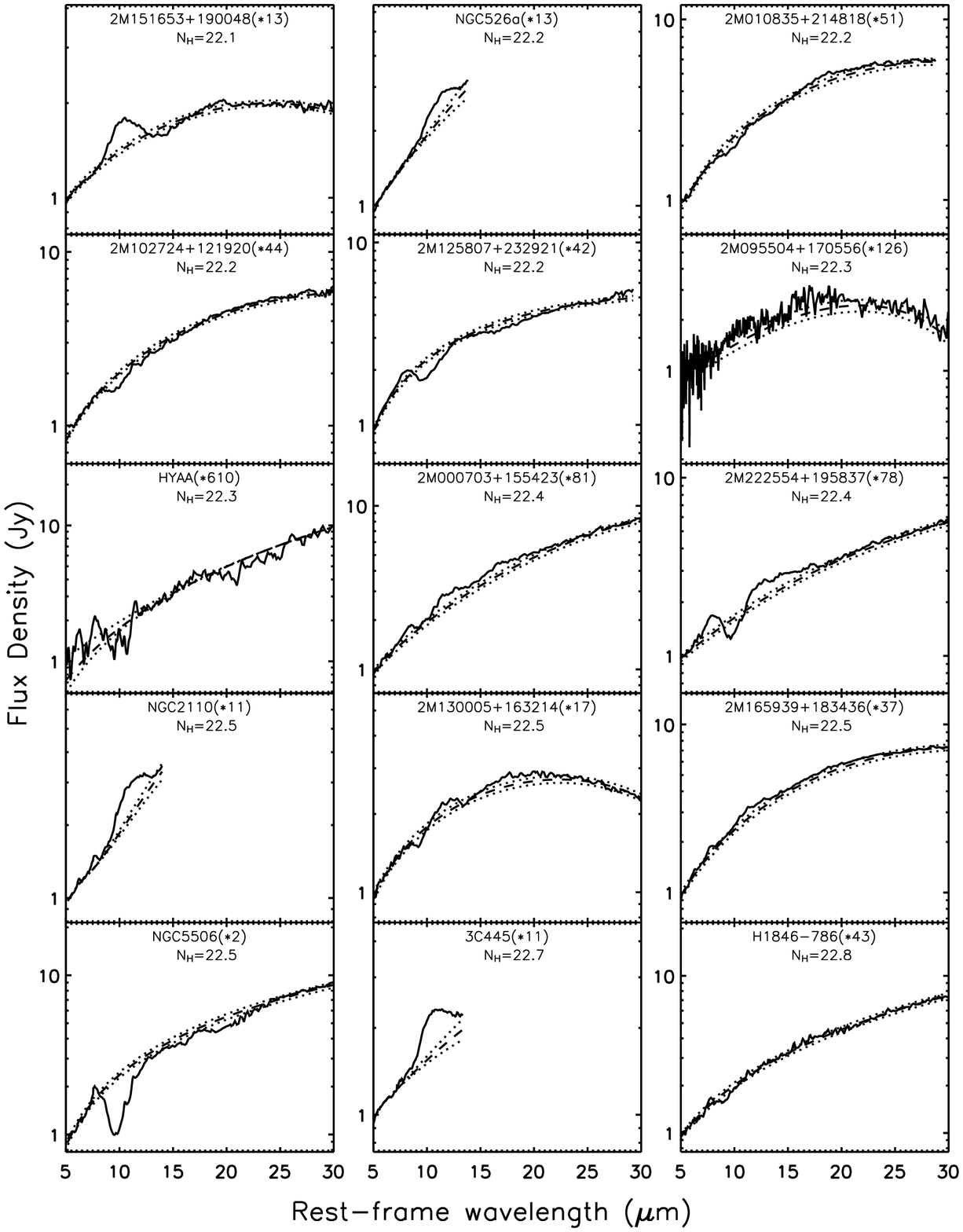}
\caption{Continued}
\end{figure}

\setcounter{figure}{1}
\begin{figure}
\epsscale{1.0}
\plotone{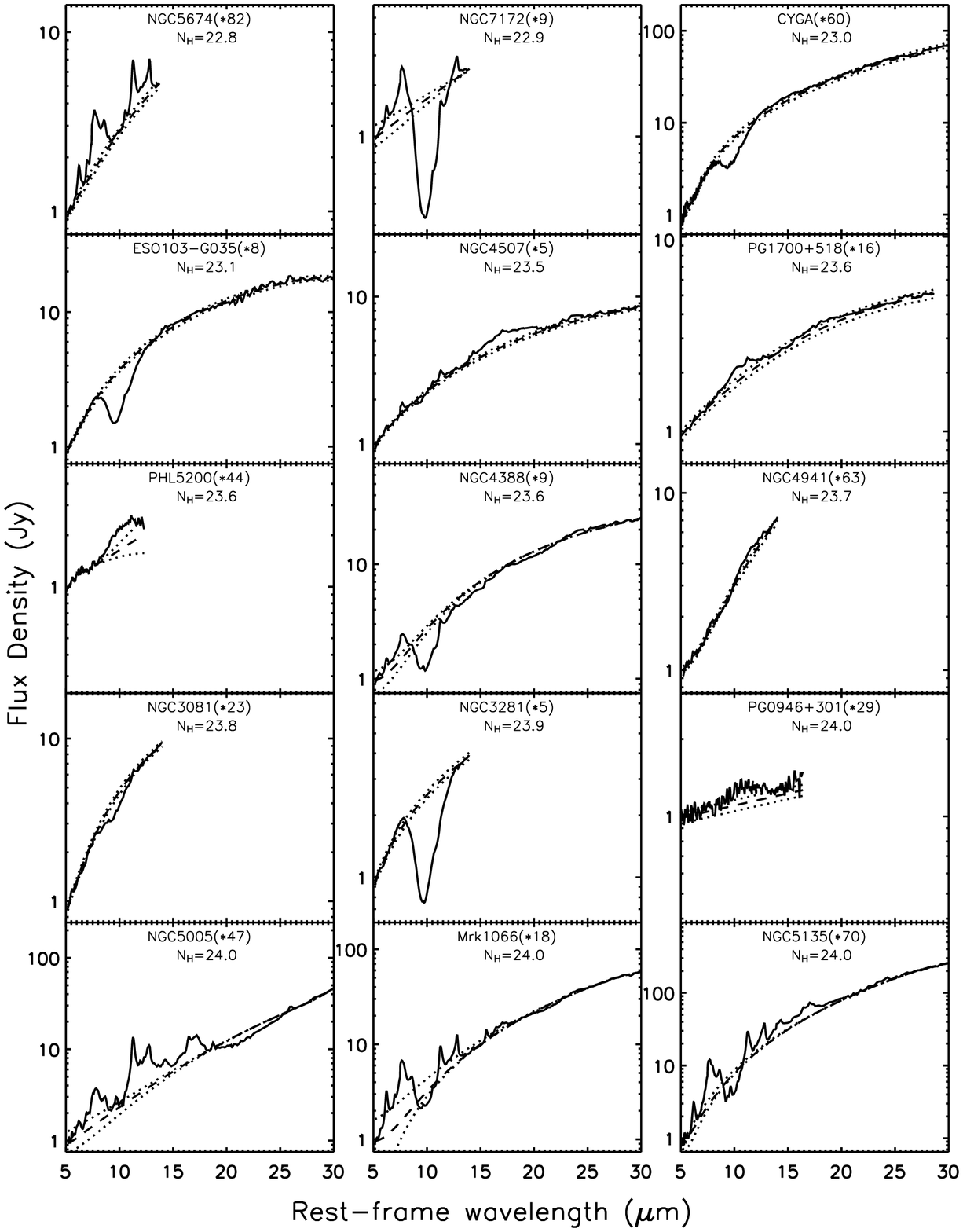}
\caption{Continued}
\end{figure}

\setcounter{figure}{1}
\begin{figure}
\epsscale{1.0}
\plotone{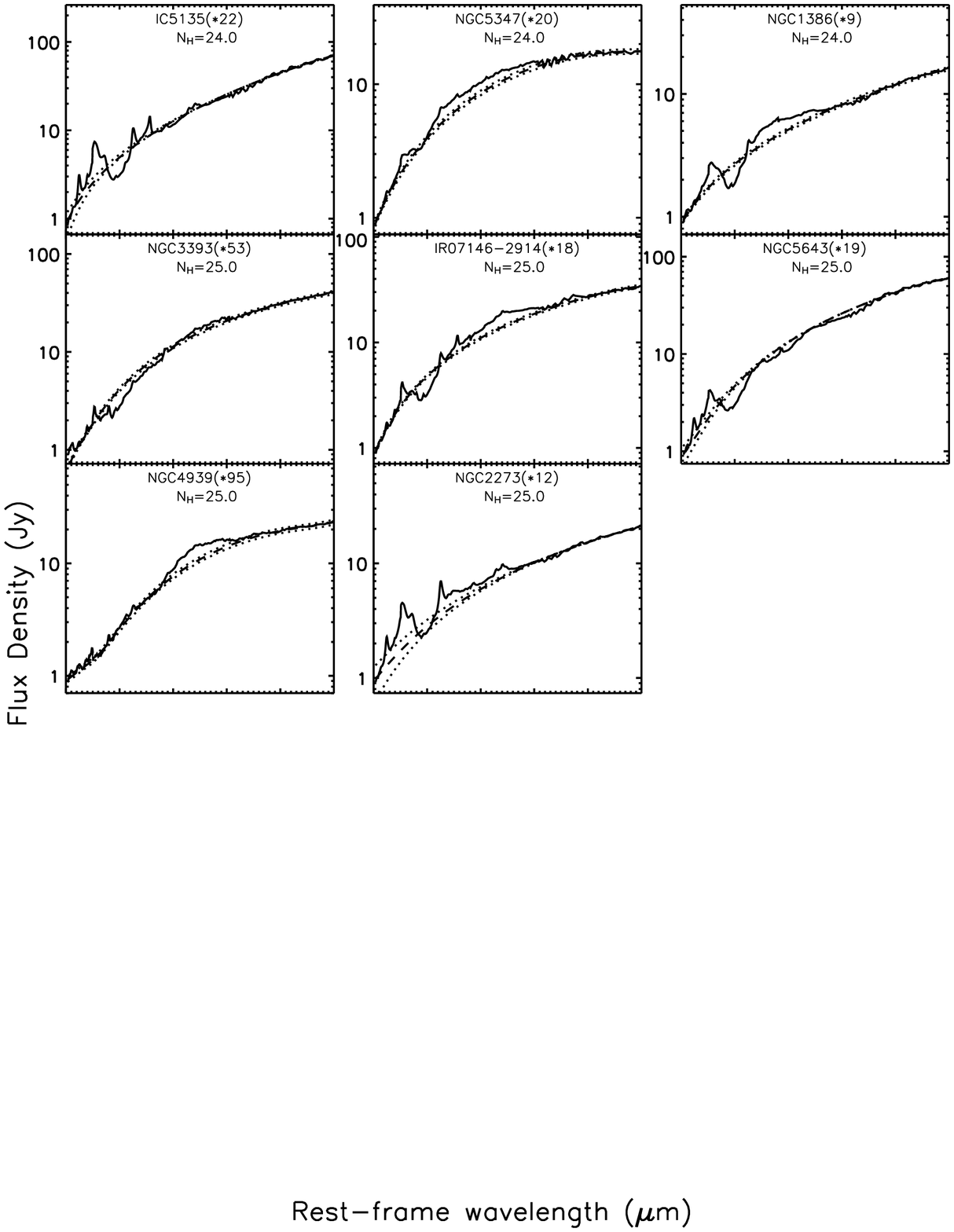}
\caption{Continued}
\end{figure}

\clearpage

\begin{figure}
\epsscale{1.}
\plotone{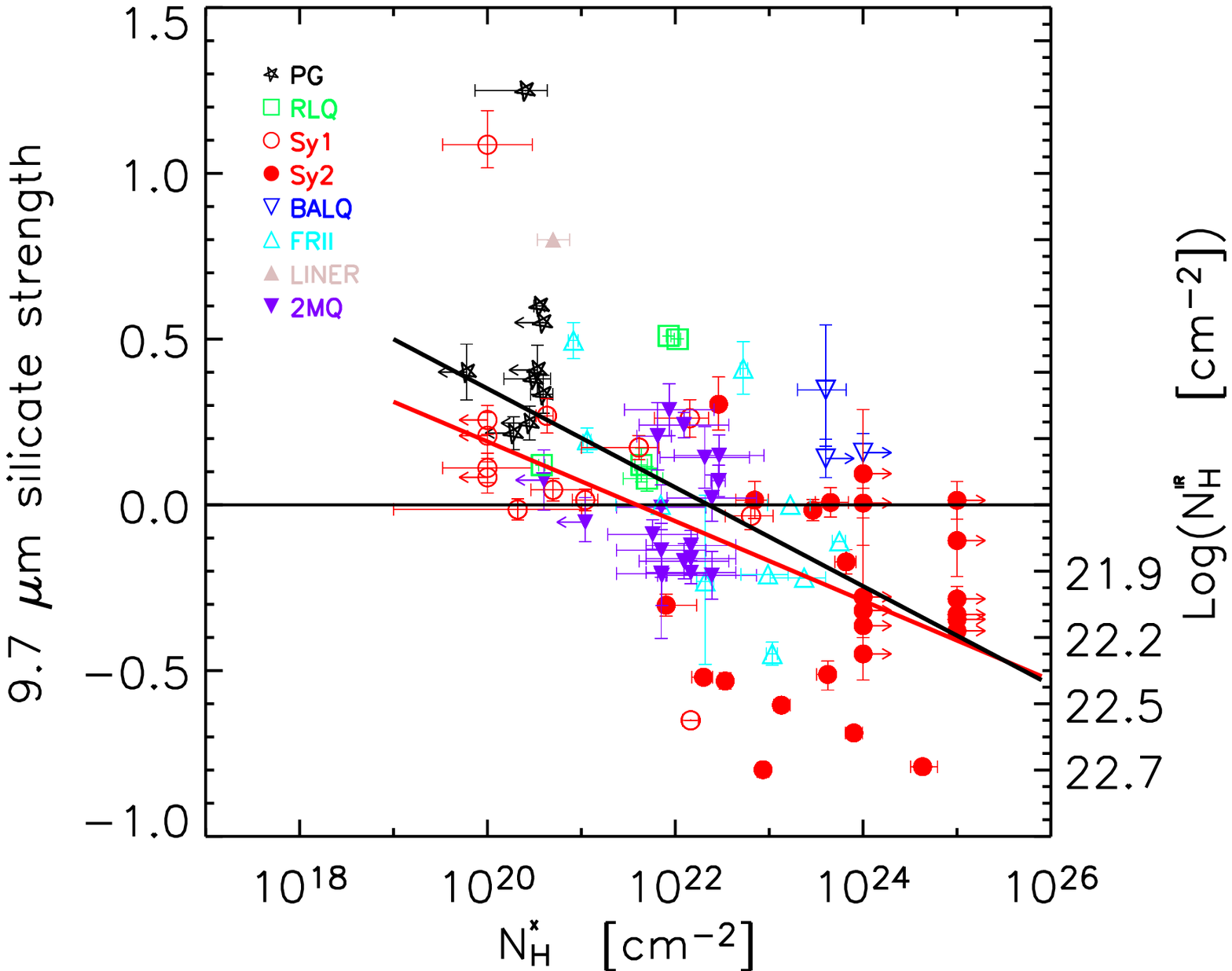} 
\caption{\label{NH_Silicate} 
The  strength of  the  silicate feature  as  a function  of HI  column density.  The strength  of the  silicate  feature is
defined as  ($F_{\rm f}-F_{\rm c})/F_{\rm  c}$, where $F_{\rm  f}$ and
$F_{\rm  c}$ are the  observed flux  density and  underlying continuum
flux density, respectively, at the  peak (for emission) or the minimum
(for  absorption) of  the silicate  feature.   The black  line is  the
linear fit  to all objects  while the red  line is the  fit to
Seyfert galaxies.  $'$PG$'$: PG quasar;  $'$RLQ$'$: radio-loud quasar;
$'$Sy1$'$:  Seyfert   1  galaxies;  $'$Sy2$'$:   Seyfert  2  galaxies;
$'$BALQ$'$:  broad  absorption-line quasar;  $'$FRII$'$:  FR II  radio
galaxies;  $'$LINER$'$: low-ionization  nuclear  emission-line region;
$'$2MQ$'$: 2MASS Quasar.}
\end{figure}

\clearpage

\begin{figure}
\epsscale{1.}
\plotone{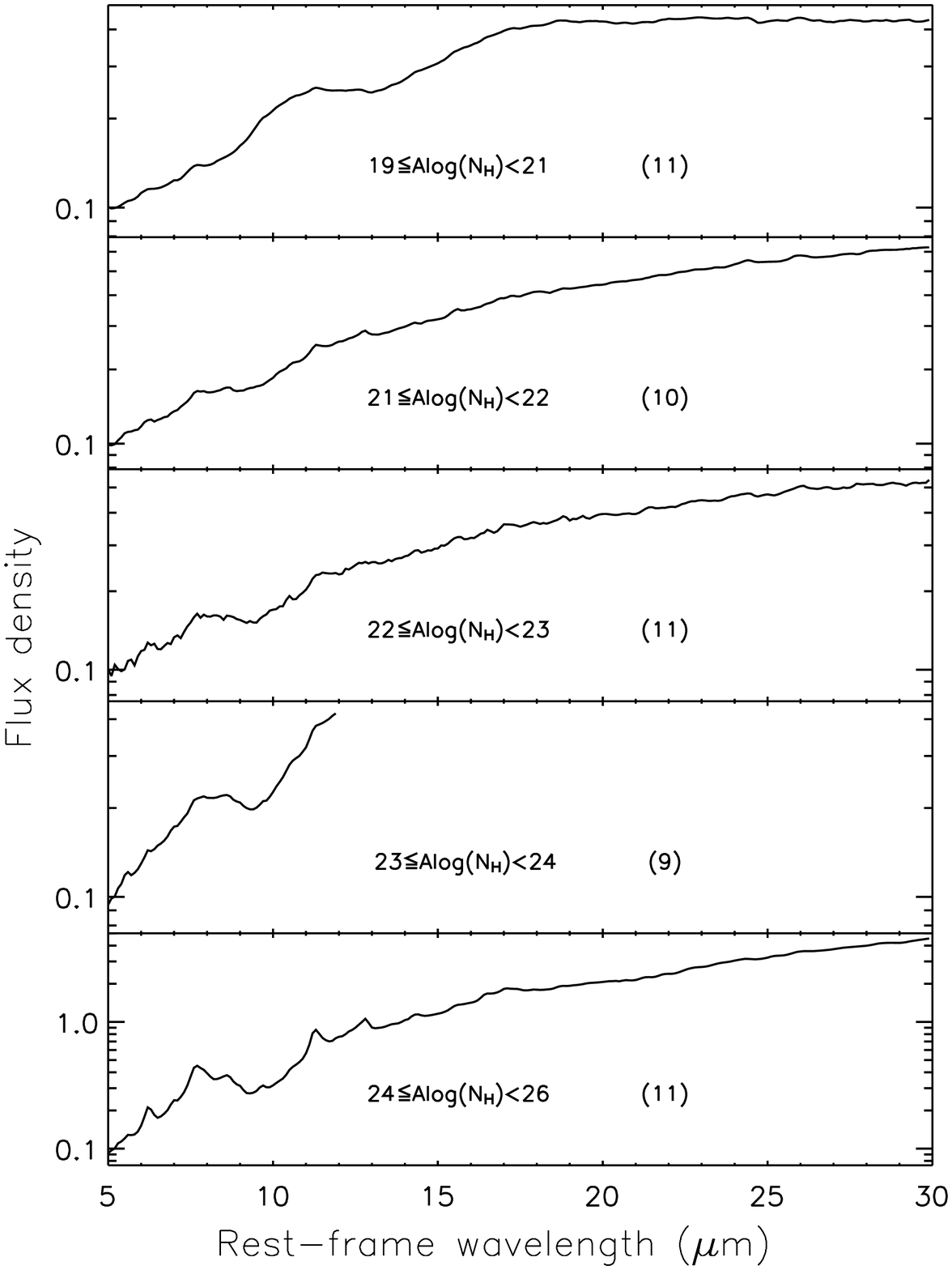} 
\caption{\label{CP_spec}  The composite spectra of AGNs in different HI column bins. 
The number in  parenthesis  is the number of objects used 
for the composite spectrum in each bin. }
\end{figure}

\clearpage

\begin{figure}
\epsscale{1.0}
\plotone{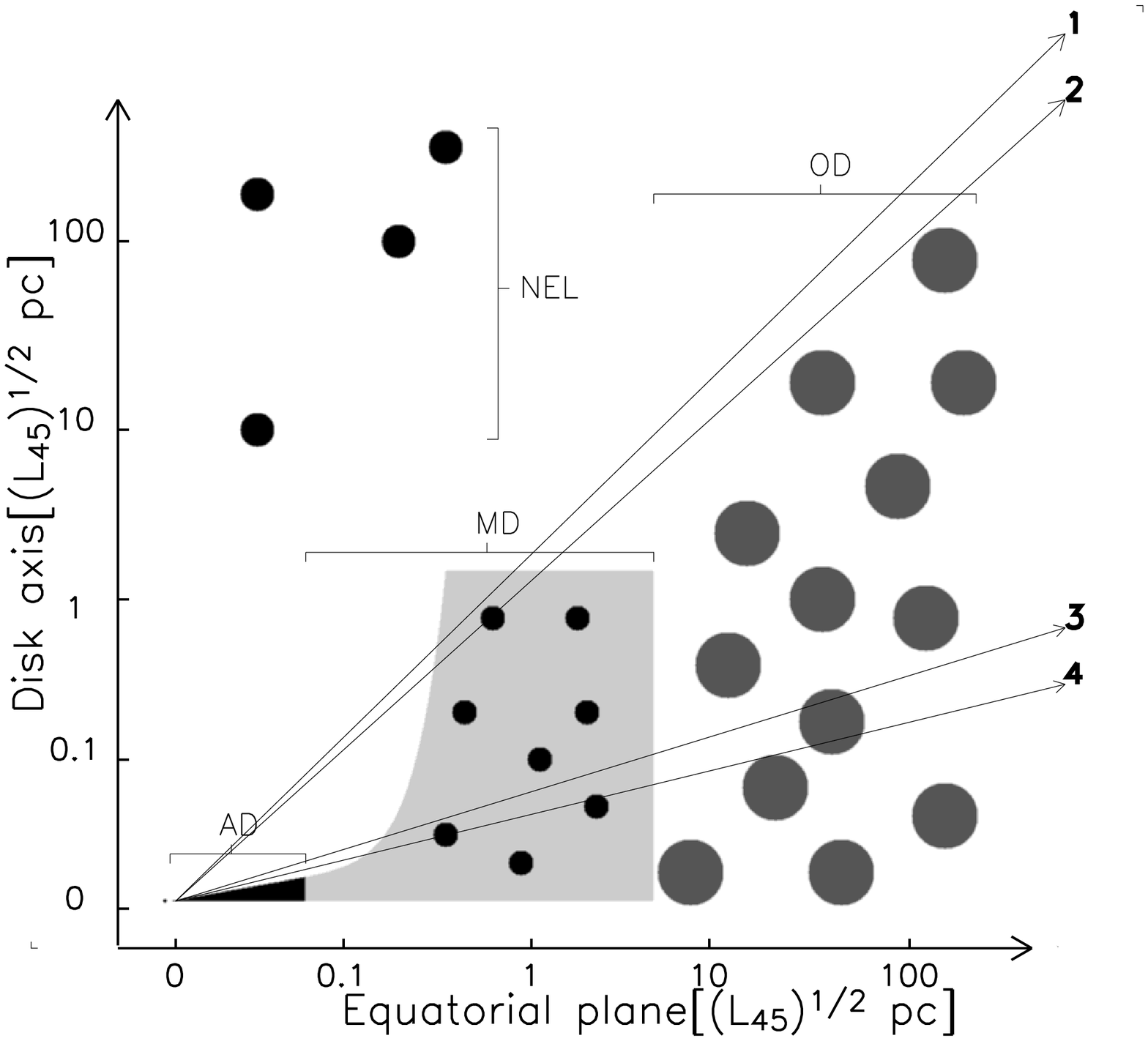}
\caption{\label{model} The  structure  of   the  material  surrounding  the  central
blackhole  in  the first  quarter  section.   The  whole structure  is
symmetric about the  disk axis and the equatorial  plane.  From inside
to outside: 1.) the inner accretion  disk (AD), which produces X-ray and UV radiation
ionizing the narrow-emission-line (NEL) and broad-emission-line
clouds,  and heating  the dust; 2.) the middle  disk (MD)  with  a diffuse
component  (grey)  and  with  denser  embedded  clouds -- the  diffuse
component  produces the  silicate emission  while the  embedded clouds
heavily obscure the central X-ray emission when the line of sight 
intercepts  them; and 3.) the  outer  disk  (OD) with  clouds  that obscure  the
silicate  and  X-ray  emission, and  are responsible  for  the  far-IR
emission. Four lines of sight indicate: 1.) silicate emission with low HI column; 2.)
silicate emission with high HI column; 3.) silicate absorption with low HI column; 
4.) silicate absorption with high HI column.}
\end{figure}

\end{document}